# Competing energy scales in topological superconducting heterostructures

Yunyi Zang[1], Felix Küster[1], Jibo Zhang[1], Defa Liu[1], Banabir Pal[1], Hakan Deniz[1], Paolo Sessi[1], Matthew J. Gilbert[2], and Stuart S.P. Parkin[1]

[1]Max Planck Institute of Microstructure Physics, Halle 06120, Germany
[2]University of Illinois at Urbana-Champaign, Department of Electrical and Computer Engineering, Urbana, IL 61820, USA

## Abstract

Artificially engineered topological superconductivity has emerged as a viable route to create Majorana modes, exotic quasiparticles which have raised great expectations for storing and manipulating information in topological quantum computational schemes. The essential ingredients for their realization are spin non-degenerate metallic states proximitized to an *s*-wave superconductor. In this context, proximity-induced superconductivity in materials with a sizable spin-orbit coupling has been heavily investigated in recent years. Although there is convincing evidence that superconductivity may indeed be induced, it has been difficult to elucidate its topological nature. In this work, we systematically engineer an artificial topological superconductor by progressively introducing superconductivity (Nb) into metals with strong spin-orbital coupling (Pt) and 3D topological surface states ($Bi_2Te_3$). Through a longitudinal study of the character of superconducting vortices within *s*-wave superconducting Nb and proximity-coupled Nb/Pt and Nb/$Bi_2Te_3$, we detect the emergence of a zero-bias peak that is directly linked to the presence of topological surface states. Supported by a detailed theoretical model, our results are rationalized in terms of competing energy trends which are found to impose an upper limit to the size of the minigap separating Majorana and trivial modes, its size being ultimately linked to fundamental materials properties.

## Introduction

In the field of condensed matter physics, Majorana fermion (MF) is an emergent fractionally-charged quasiparticle that obeys non-Abelian exchange statistics (1,2), and has been endlessly purported to be the foundation of topological quantum computation (3). Generally speaking, MF are expected to appear in the core of vortices proliferated in superconducting condensates with *p*-wave pairing symmetry (4). A promising route to realize MFs relies on the creation of topological superconductor heterostructures where spin-split metallic states are proximitized to an *s*-wave superconductor (SC) (5). In this context, theoretical predictions suggest that the proximity effect between an ordinary *s*-wave superconductor and the Dirac surface states of a 3D time-reversal invariant topological insulator may lead to the emergence of MF within vortices (6). A similar realization scheme has been applied to ordinary electron systems characterized by strong spin-orbit coupling (SOC) and a large *g*-factor (7,8). Under these circumstances, a spin non-degenerate 2D electron gas is similar to the surface states of 3D topological insulators (TI) may be obtained by considering the combination of SOC and Zeeman effect. The SOC splits the spin degenerate band into a pair of spin non-degenerate bands while the Zeeman effect opens a gap at the crossing point of these two bands. When the chemical potential is tuned into this Zeeman gap, there is only one Fermi surface with helical spin polarization. A drawback of the above scheme is that the size of the Zeeman splitting must be larger than the size of the induced superconducting gap, a condition difficult to meet in ordinary 2D metals due to the orbital pair-breaking effect (67).

In the case of TIs, the Dirac, spin-split bands on the surface allows to avoid the complications related to the presence of degenerate time-reversed pairs seen in the strongly SOC systems. Proximity-induced superconductivity in prototypical 3D TIs such as $Bi_2Se_3$ and $Bi_2Te_3$ has been studied by scanning tunneling microscopy (STM), angle-resolved photoemission spectroscopy (ARPES) and quantum transport measurements each of which reveal indirect yet tantalizing glimpses indicating the presence of MFs (11,14-16,59). However, the topological insulators films used in the overwhelming majority of previous works are heavily electron doped, with a Fermi level lying well inside the bulk conduction bands. In such a scenario, topologically trivial and non-trivial states coexist, complicating the interpretation of the experimental results. Differences in the resultant manifestations of topological superconductivity that result from the physical properties inherent to different substrates are still not understood. For example, Cooper pairing in the Dirac surfaces states has been reported for $Bi_2Se_3$ grown on Nb (16) and for $Bi_2Se_3$ and $Bi_2Te_3$ on $NbSe_2$ substrates (13,16). On the other hand, absence of proximity-induced gaps have been reported for $Bi_2Se_3$ coupled to optimally-doped cuprate superconductors (61), a result in sharp contrast to fully gapped surfaces states reported in an earlier study (11). Several factors such as interface quality, superconducting penetration length, the presence of interface states, and interfacial lattice mismatch have been invoked to explain these different results (17–21). More recently, the fabrication of bulk-insulating $(Bi_xSb_{1-x})_2Te_3$/Nb heterostructures (x = 0.62) by flip-chip technique (10) resulted in the absence of proximity induced superconductivity even for $(Bi_xSb_{1-x})_2Te_3$ films only 2 layers thick (10). These results, compared to heavily doped $Bi_2Se_3$ films grown on the same substrate, suggested that the bulk states play a crucial role in transiting superconductivity to the topological Dirac states (10).

Additionally, the unambiguous detection of MF signatures is severely complicated by the presence of Caroli-de-Gennes-Matricon (CdGM) states, low-energy excitations emerging within vortex cores of type-II superconductors which are characterized by a discrete energy spectrum with the lowest state emerging at about $E = \pm\Delta^2/E_F$, with $\Delta$ being the superconducting energy gap and $E_F$ the Fermi level (68). Due to the very-small value of $\Delta/E_F$, CdGM states are generally detected as a symmetric peak in the local density of states centered at zero energy (69).

In this work, we report a longitudinal study that examines the materials issues that are endemic to the observation of MF in vortex cores of TIs by fabricating an artificial topological superconductor. Starting from superconducting Nb (110) films showing "clean" vortices, i.e. vortices without Caroli-de-Gennes-Matricon state, we progressively introduce strong spin-orbit coupling and topological states by proximitizing Pt and, finally, bulk-insulating $Bi_2Te_3$ films, respectively. By directly comparing the detailed spectroscopic characterization performed on all heterostructures, we reveal materials-dependent signatures through which MFs emerge only in the $Bi_2Te_3$ case. Backed up by theoretical simulations, our results provide compelling experimental evidence that the details of the underlying TI are not the impediment to the clear observation of topological superconductivity hosting Majorana modes but rather due to the existence of competing energy trends directly linked to fundamental materials properties and that set an upper limit to the maximum size of the induced minigap in the vortices.

**Result**

*Fabrication and characterization of topological superconducting heterostructures*:

In Figure 1 we illustrate the cross-sectional STEM and STM images of the different heterostructures that we consider in this work. We begin our examination of the materials properties that lead to stable topological superconductivity by considering the vortex states in niobium (Nb). In Figure 1a, we show a cross-sectional scanning transmission electron microscopy (STEM) image of the $Nb/Al_2O_3$ heterostructure. Niobium represents an optimal *s*-wave superconducting material offering several advantages: (i) it has the highest transition temperature among all elemental superconductors (T = 9.2 K) and (ii) being a type-II superconductor with formation of vortex in magnetic field, it affords the option of studying well-formed superconducting vortices as a baseline for later heterostructures formed with niobium. *(3,4,6).}*

Nb films of thickness in the range of 4-7 nm are deposited on $Al_2O_3$ (see Supplementary Information for a description of sample preparation). The sharp interface and clear atomic resolution highlights the crystalline quality of the Nb films. Figure 1b shows a topographic STM image acquired in the constant current mode. The surface consists of large terraces with few islands on it, which are also indicative of the high quality of our films. As shown in the inset, the step height matches the monoatomic distance between subsequent atomically flat planes along the Nb (110) crystallographic direction. This low surface roughness is crucial to create sharp interfaces and promotes a better epitaxy of the films subsequently grown on top of it (28, 29), as described in the following.

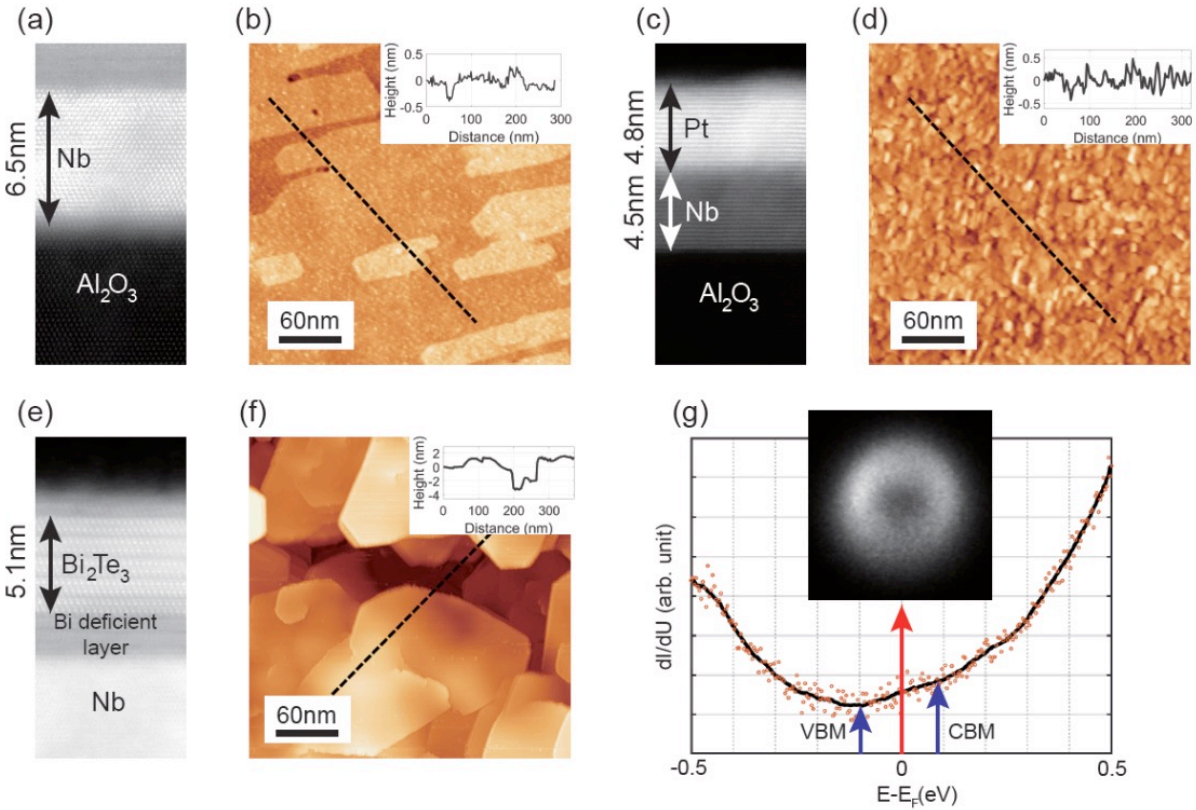

**Figure 1: The heterostructures lineup** (a, b) Nb/Al$_2$O$_3$, (c, d) Pt/Nb/Al$_2$O$_3$, and (e,f) Bi$_2$Te$_3$/Nb/Al$_2$O$_3$ heterostructures. For each heterostructure, the (a, c, e) and (b, d, f) panels report a STEM cross-sectional and STM surface image of the samples, respectively. (g) Spectroscopic characterization of the Bi$_2$Te$_3$ film. Both STS and ARPES data (inset) demonstrate that the Fermi level lies well-inside the bulk gap, where only topological surface states exist. VBM and CBM refer to valence band maximum and conduction band minimum, respectively.

To create a superconducting condensate in a material characterized by strong SOC, a thin Pt film has been directly deposited onto Nb to induce superconductivity by the proximity effect. A cross-sectional TEM image of the resulting heterostructure is reported in Figure 1c, demonstrating the creation of a sharp Pt-Nb interface, an important aspect to allow Cooper pairs to efficiently tunnel into the Pt film (35). The Pt layer grows with crystalline quality along the (111) direction and has a thickness of 4.8 nm. The surface topography (Figure 1d) shows a homogeneous Pt film characterized by low surface roughness.

Finally, to scrutinize the effect of a topologically non-trivial band structure on proximity-induced superconductivity, thin films of the prototypical topological insulators Bi$_2$Te$_3$ have been grown onto the same Nb underlayers. It is worth stressing that, contrary to previous reports where unconventional superconductors such as BSCCO (11) or NbSe$_2$ (12–15) are used, our Bi$_2$Te$_3$/Nb heterostructure represents the easiest possible platform for engineering topological superconductivity by directly coupling a TI to a conventional s-wave SC (3,4,6). The (110) orientation of the underlying Nb film is found to promote the growth of $Bi_2Te_3$ along the (111) crystallographic direction (see RHEED patterns reported in the Supplementary Information). The thickness of our films amounts to 5 quintuple layers (QL) of the $Bi_2Te_3$ crystal structure, with each QL layer corresponding to the sequence Te-Bi-Te-Bi-Te (26). The aforementioned sequence is clearly discernible in the STEM image reported in Figure 1e. 5 QL thickness maximizes the strength of the proximity effect while at the same time avoiding strong hybridization effects between the Dirac states hosted on top and bottom surfaces, which is characteristic of thin-film TI systems (71). The surface morphology visualized by STM consists of atomically flat terraces (Figure 1f). The line profile (inset of Figure 1f) reveals steps corresponding not only to the expected QL height, but also to fractional-QL heights. As observed in earlier studies, this is a direct influence of the underlying substrate surface morphology and reactivity, which imposes a vertical translation between two adjacent domains (36). However, as described in the supporting information, this is found to do not have any influence on the superconducting properties, which are homogeneous across the entire sample.

Being prototypical TIs narrow gap semiconductors (37), a precise spectroscopic characterization is crucial to precisely locate the Fermi level with respect to the bulk valence and conduction bands. Indeed, doping effects have been shown to shift the Fermi level well-inside the bulk bands already at moderate defect concentrations (38–41), a scenario highly unfavorable to an unambiguous identification of topological effects. Scanning tunneling spectroscopy (STS) data reported in Figure 1g demonstrate that our $Bi_2Te_3$ films have a Fermi level residing well-inside the bulk gap located at approximately the middle of the gap defined by the valence band maximum (VBM) and conduction band minimum (CBM) (see Ref. (42, 43) for a detailed description of energy level positioning). The position of the Fermi level is further corroborated by the constant energy cut obtained by ARPES at room temperature (reported in the inset of Figure 1g), showing the typical isotropic shape of Dirac-like topological states and the absence of bulk bands. These observations make our TI/SC heterostructures ideal platforms to investigate topological superconductivity, since in sharp contrast to similar studies focusing on other prototypical TIs such as $Bi_2Se_3$ (heavily n-doped, (38, 39)) or $Sb_2Te_3$ (heavily p-doped (40, 41)), the absence of trivial states lying at the Fermi level allows to determine the impact of material and interface quality as key ingredients required to engineer topological superconductivity (3,4,6,26).

## Superconducting gaps induced by proximity effect

The superconducting properties of the heterostructures have been investigated by STS measurements performed at cryogenic temperatures. Figure 2 reports a series of differential conductance dI/dU spectra obtained at progressively higher temperatures on all systems investigated in the present study, namely: superconducting Nb (Figure 2a), proximity induced superconductivity in a strong spin-orbit coupled material Pt/Nb (Figure 2b), and proximity induced superconductivity in a topologically non-trivial material: $Bi_2Te_3$/Nb (Figure 2c). A clear superconducting gap is visible in all systems at the lowest temperature achievable in our set-up, i.e. 500 mK. However, both quantitative as well as qualitative differences exist among the heterostructures. While Nb films show a superconducting transition temperature of approximately 9 K, in agreement with bulk data (44), a significantly lower temperature is necessary to create a superconducting condensate in Pt/Nb and Bi$_2$Te$_3$/Nb heterostructures. In both systems, superconductivity emerges only below 4 K, the lower transition temperature being a direct consequence of superconductivity induced by the proximity effect. Additionally, while the Nb and Pt/Nb systems are both well-fitted by an *s*-wave BCS-type spectral function (45, 46), this not true for Bi$_2$Te$_3$/Nb. A strong deviation from standard *s*-wave BCS behavior is clearly signaled by the very sharp single particle coherence peaks visible at the boundary of the excitation gap (12,14). The use of the same underlayer (Nb) for both Pt and Bi$_2$Te$_3$ films, their similar thickness, and their same superconducting transition temperature set the foundation for a meaningful comparison in terms of their different electronic structures. In particular, our data serve to clearly highlight the decisive role played by the properties of the TI in the observation of phenomena pertaining to unconventional superconductivity under optimal conditions, i.e. the only states existing at the Fermi level in our Bi$_2$Te$_3$ films are those associated with the topological bands.

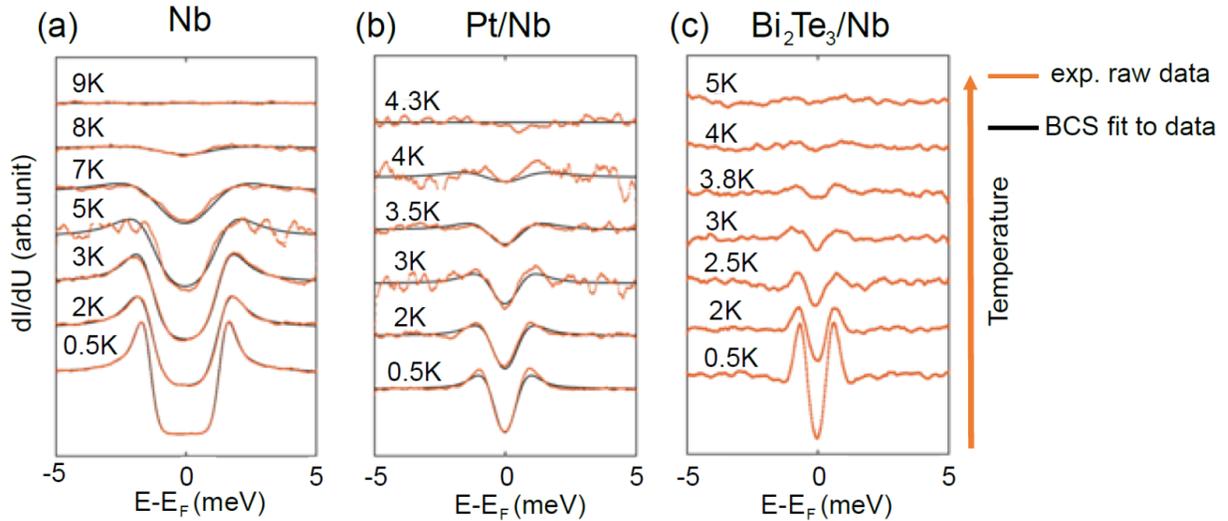

**Figure 2: Temperature-dependent scanning tunneling spectroscopy** Superconducting energy gap observed in (a) Nb/Al$_2$O$_3$, (b) Pt/Nb/Al$_2$O$_3$, and (c) Bi$_2$Te$_3$/Nb/Al$_2$O$_3$ heterostructures at progressively higher temperatures. The experimental data (orange line) can be reproduced by a BCS fitting (black line) for Nb/Al$_2$O$_3$ and Pt/Nb/Al$_2$O$_3$. A strong deviation from this behaviour is clearly visible for Bi$_2$Te$_3$/Nb/Al$_2$O$_3$.

## Spectroscopic mapping of vortices

The different Cooper pairing mechanisms among the heterostructures are schematically illustrated in Figure 3 a,b, and c for Nb, Pt/Nb, and Bi$_2$Te$_3$/Nb, respectively. Without any significant spin-orbit coupling (Nb), the Fermi surface consists of two spin-degenerate energy bands. The pairing between electrons (indicated by the dashed line) can be described by the conventional BCS theory and the resulting Cooper pairs are characterized by a standard *s*-wave singlet state. By introducing a heavy element thin film (Pt), the two spin bands are split in momentum space because of the combined action of spin-orbit coupling and lack of out-of-plane inversion symmetry naturally occurring at surfaces and interfaces, an effect known as Rashba effect (62,63). The introduction of SOC to the heterostructure opens the possibility to create more complicated and potentially unconventional superconducting phases. In particular, the presence of strong SOC and magnetic fields may effectively introduce a *p*-wave superconducting pairing ($p_x \pm ip_y$) in addition to the conventional *s*-wave, which is known to play a crucial in the formation of topologically non-trivial superfluids hosting Majorana modes. However, since fermion states still occur in degenerate time-reversed pairs in a Nb/Pt heterostructure, turning a spin-split 2D metal into a topological superconductor requires the introduction of a Zeeman terms to imbalance the two *p*-wave degenerate components, resulting in an effective $p_x + ip_y$ superconductor. The degeneracy splitting induced by the magnetization must be larger than the size of the induced superconducting gap (64-67), a condition difficult to meet in ordinary metals due to the orbital pair-breaking effect. As illustrated in Figure 3c, the odd number of spin-split bands hosted on the surface of topological insulators represent an ideal solution to induce topological superconductivity where Majorana fermions are predicted to appear as zero-energy bound states in superconducting vortex cores.

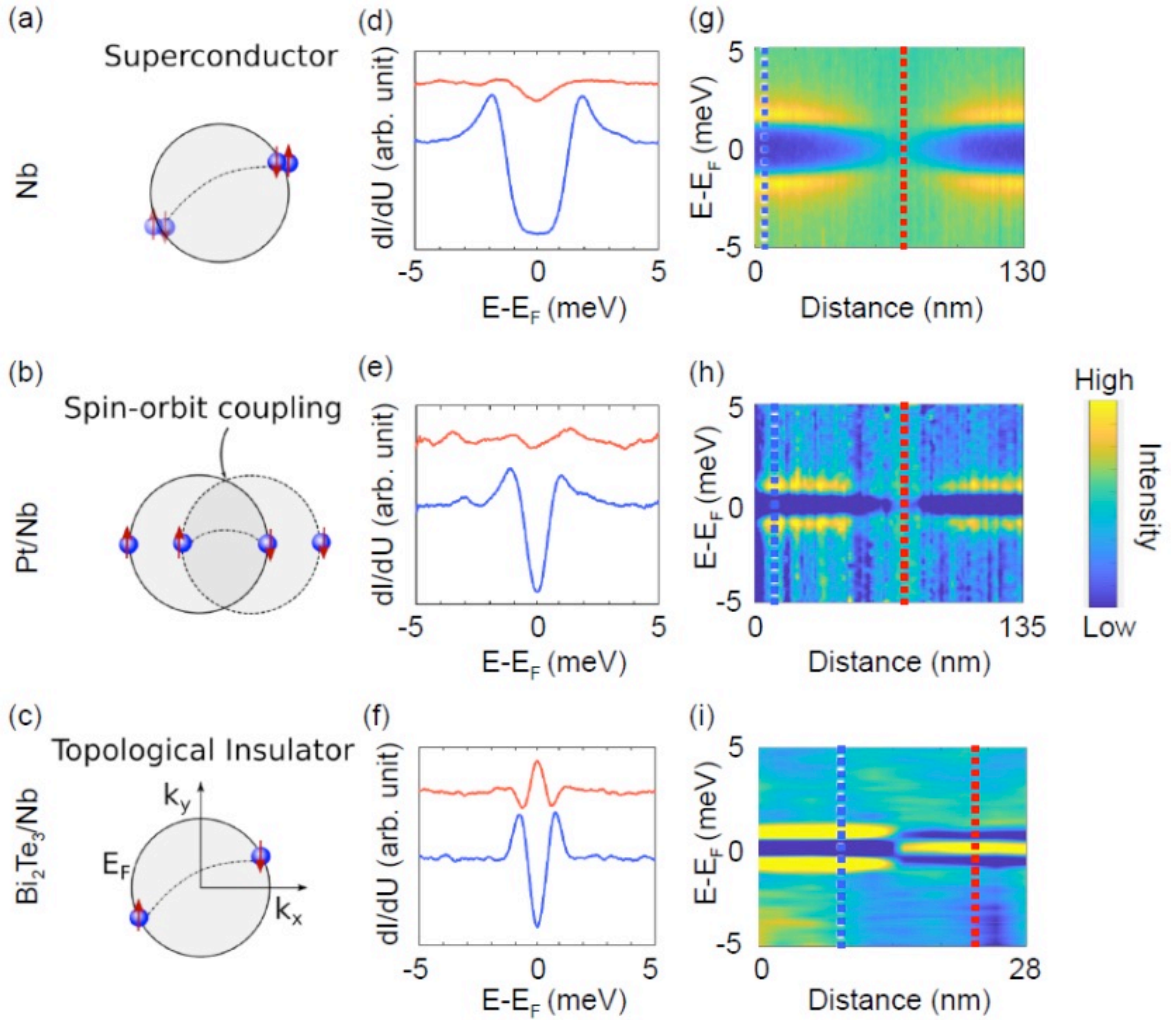

**Figure 3: Spectroscopy across vortices** (a-c) Schematic illustration of superconducting pairing, (d-f) scanning tunneling spectroscopy data acquired by positioning the tip far away from the vortex (blue line) and at the vortex core (orange line), (g-i) spectroscopic profile across a vortex taken, for each heterostructure, along the white lines visible in the dI/dU maps in Figure S2. For each heterostructure, the vertical lines visible in (g-i) correspond to the positions where the spectra reported in panels (d-f) have been acquired.

The impact of the different electronic and spin-textures onto the properties of the superconducting condensates has been experimentally probed by spectroscopic measurements acquired with magnetic fields applied perpendicular to the sample surface. In agreement with expectations for type-II superconductors, superconducting vortices appear in all heterostructures, i.e. Nb, Pt/Nb, and $Bi_2Te_3$/Nb, with the vortex density progressively increasing with increasing magnetic field strength (see Figure S4 in the Supplementary Information). However, unambiguous spectroscopic differences among the heterostructure are clearly revealed in Figure 3, with panels d, e, and f reporting STS spectra taken by positioning the tip in the vortex core (red lines) and far away from it (blue lines) for Nb, Pt/Nb, and $Bi_2Te_3$/Nb, respectively. While the superconducting gap vanishes at the vortex core in both Nb and Pt/Nb, a strong zero-bias peak emerges in the $Bi_2Te_3$/Nb case. A direct comparison of the data obtained on the three different heterostructures confirms the crucial role of the topological states, as opposed to simply strong SOC, in determining the properties of the $Bi_2Te_3$/Nb superconducting condensate.  In particular, the zero-bias peak detected in the $Bi_2Te_3$/Nb case (see Fig. 3 i) is consistent with the theoretically predicted signature of Majorana modes, which are expected to emerge in the vortex core of topological superconductors (6). It is worth stressing that type-II superconductors are known to host topologically trivial mini-gap bound states in their vortex core, i.e. the so-called Caroli-de-Gennes-Matricon states (48). However, their absence in the Nb film suggests that their contribution is strongly reduced in our case, contrary to an earlier study where strong trivial Caroli-de-Gennes-Matricon where present in the substrate (16). Additional experimental support for Majorana modes come from the analysis of the spatial evolution of the spectroscopic properties in applied magnetic fields. Figure 3g, h, and i report line spectra acquired across the superconducting vortices for Nb, Pt/Nb, and $Bi_2Te_3$/Nb respectively (measured along the white lines visible in their respective dI/dU maps, see Supplementary Figure 4). For Nb and Pt/Nb, the single particle coherence peaks become progressively weaker while simultaneously converging into an X-shaped feature whose crossing point is located at the vortex core. On the other hand, a sharp transition to a zero-bias peak is evidenced in the $Bi_2Te_3$/Nb case. In agreement with theoretical predictions for Majorana modes and in sharp contrast to the typical spatial behavior of Caroli-de-Gennes-Matricon states, the zero-bias peak does not split right off the vortex center, showing constant intensity over tens of nm (49, 50).

## Modeling topological superconducting heterostructures

To understand the physics contained within the vortex in proximity-coupled s-wave SC-3D time-reversal invariant topological insulators (STI) system, we diagonalize a 3D TI Hamiltonian proximity-coupled to an s-wave superconductor (see Supplementary Information for detail). In our model, we consider two different cases of the penetration depth ($\lambda$) of magnetic field: the "thin-flux" limit and the "thick-flux" limit (51). Specifically, the "thin-flux" limit occurs when the topological insulator layer is sufficiently thin and the magnetic flux that penetrates into the bottom layer of the topological insulator does not have sufficient distance to spread before reaching the top surface. Within the parameters that we have defined for the STI system, the "thin-flux" limit occurs when $\lambda \approx a$, with a being the lattice constant. In the "thick-flux" limit, that occurs when $\lambda \gg a$, the penetrating flux is dispersed uniformly within the TI film prior to reaching the top surface. Figure 4a reports the energies of the two lowest energy excitations emerging in the STI heterostructure as a function of the penetration depth $\lambda$, spanning from the "thin-flux" to the "thick-flux" regime. These states correspond to the Majorana fermion (black line) and the first trivial bound state (red line), which are both localized within the vortex core. While the energy of the Majorana fermion approaches zero by progressively increasing $\lambda$, the topologically trivial state remains constant in energy as the penetration depth is changed. Their energy separation corresponds to the size of the minigap, which at first becomes larger by increasing $\lambda$, and then stays constant at its maximum value reached at $\lambda=1$ (expressed in units of the lattice constant a).

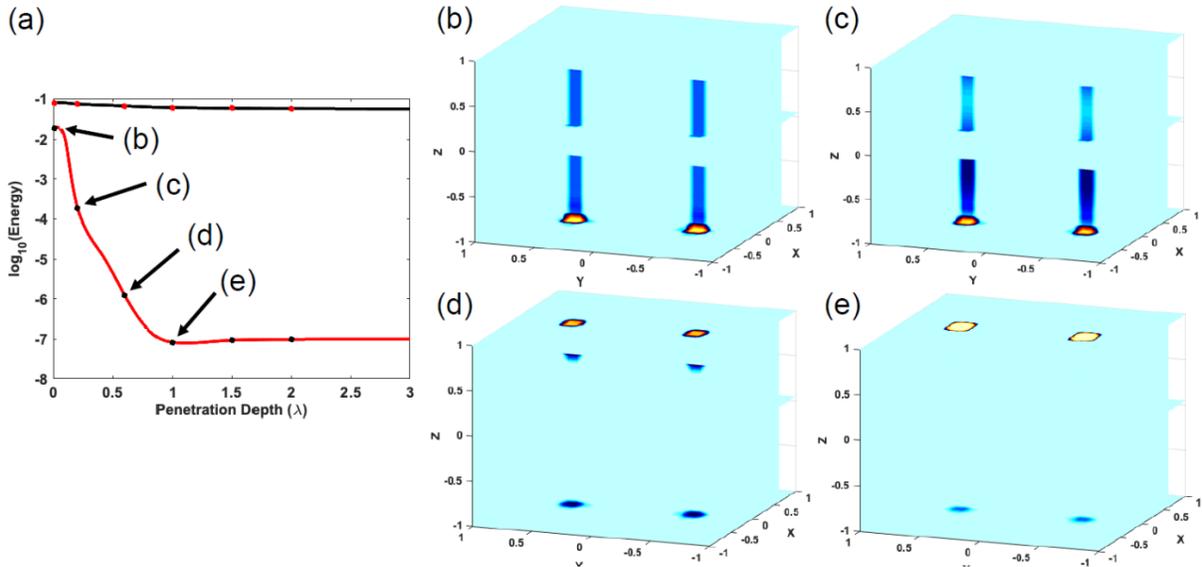

**Figure 4: Probability Distribution of Lowest Energy States in 3D**: (a) Plot of the energy for the lowest and second lowest energy modes in the 3D system for $\lambda = 0.2$. We observe the formation of the topological phase in which the energy of the lowest lying state, the Majorana fermion localized at the vortex core, approaches zero with increasing $\lambda$ while the trivial second lowest state remains constant in energy as the penetration depth is changed. The probability density distributions corresponding to penetrations depths of (b) $\lambda = 0.01a$ (c) $\lambda = 0.1a$ (d) $\lambda = 0.5a$ and (e) $\lambda = 1.0a$ showing the evolution of the states from being delocalized along the magnetic flux tube to localized in the vortex cores on the surface as the penetration depth of the superconductivity is increased.

Figure 4b-e illustrates the spatial distribution of the lowest lying state in the STI heterostructure, corresponding to penetrations depths of (b) λ = 0.01a (c) λ = 0.1a (d) λ = 0.5a and (e) λ = 1.0a. Within the "thin-flux" limit (panel b), the physics is dominated by wormhole Majorana states largely delocalized into the bulk such that they can tunnel through the vortex and hybridize, opening an energy gap which pushes them closer to the topologically trivial low state, as illustrated in Figure 4a. Figures 4c and d report the results obtained for λ values corresponding to the transition between the "thin-flux" and the "thick-flux" regimes, showing how Majorana states become progressively more localized onto the surface as the penetration depth is increased, reaching a scenario where the low energy mode if completely localized onto the surface for λ = 1 (panel). This regime corresponds to the best possible scenario for the detection of MFs in tunneling experiments, since it allows to simultaneously maximize both their spectral weight onto the surface as well as the size of the minigap which separates them from topologically trivial states.

## Discussion

The "thick-flux" condition is obviously fulfilled in our $Bi_2Te_3$/Nb samples, being that the thickness of the TI film (~ 5 nm) is much smaller than the Nb penetration depth (~ 40nm (70)), creating the best possible experimental conditions for the detection of Majorana modes. In our system, the zero-bias peak emerging inside vortices may be directly linked to the topological states, which are the only states present at the Fermi level in our bulk-insulating $Bi_2Te_3$ films. The spatial mapping is also consistent with theoretical predictions suggesting a local density of states distribution of Majorana modes that resembles a Y shape (see Fig. 3i) which is qualitatively distinct from the typical V-shape of CdGM states (69). However, the small energy gap separating MFs from CdGM states represents the major obstacle to energetically distinguish MFs from trivial states. Indeed, as evidenced by our model, the minigap size reaches a maximum which is directly related to fundamental material properties such as the band structure. With a proximity-induced SC gap of approximately 1 meV (see Fig. 2 c) and a Fermi level located 150 meV above the Dirac point (see Fig. 1 g), the minigap separating MFs from trivial states has a size of 0.01 meV. This value imposes severe experimental conditions to reach the quantum limit situation $T/T_c \ll \Delta/E_F$ which would allow to establish a direct relation between zero energy modes and MFs. A larger $\Delta/E_F$ value might be possible by either increasing the SC gap or by lowering the Fermi level. However, for a given class of materials, these values may not be independently changed, being directly linked to each other by competing and opposite energetic trends. For example, moving the Fermi level $E_F$ closer to the Dirac point to increase the $\Delta/E_F$ ratio is intrinsically accompanied by a reduction of the proximity-induced superconducting energy gap $\Delta$, the smaller gap being a direct consequence of the lower density of states at the Fermi. Being that these effects are intrinsically tied to one another, our results reveal that the size of the minigap, and consequently the ability to energetically resolve MFs, is ultimately limited by fundamental and intertwined material properties.

## Conclusion

The emergence of proximity-induced superconductivity in bulk-insulating $Bi_2Te_3$ demonstrates that topological Dirac states can be effectively driven into Cooper pairs. Our results provide compelling experimental evidence that the creation of a superconducting condensate onto the surface of TIs does not require the presence of bulk states (10), in but does depend on the interface properties between materials, location of the Fermi energy, the induced superconducting gap, and the topological band structure. Our spectroscopic measurements reveal a strong deviation from a conventional BCS spectrum which can be directly linked to the presence of the topological states, as evidenced by the comparison with the Nb and the Pt/Nb cases (16). Additionally, we observe a zero-bias peak that emerges inside superconducting vortices in the $Bi_2Te_3$/Nb. The energy position and spatial distribution are found consistent with the expected signatures for Majorana modes (50). Based on a detailed theoretical model, our results are rationalized in terms of the minigap separating MFs from CdGM states, revealing that the energetic separation between states may be maximized by driving the system into the thick flux regime. However, the existence of competing energy trends provides an impediment to arbitrarily increasing its size beyond a maximum value dictated by fundamental materials properties. Overall, our results unveils the mechanisms on which is necessary to act to strengthen the proximity effects in TIs-Nb, evidencing the existence of inherent material limitations which underlie the unambiguous detection of Majorana modes in topological superconductors.